\documentclass[aps, prl, reprint, amsmath, amssymb, 10pt, showkeys, onecolumn]{revtex4-2}
\usepackage{graphicx} 
\usepackage{dcolumn}  
\usepackage{bm}     
\usepackage{subcaption}
\begin{document}

\title{Acrylamide Conformers: A Revision of Published Density Functional Theory Studies}

\author{William Scott}
\affiliation{Center for Simulation and Modeling, George Mason University, and
Department of Computational and Data Sciences, George Mason University,
Fairfax, Virginia, United States}
\author{Estela Blaisten-Barojas}   
\email{blaisten@gmu.edu}
\affiliation{Center for Simulation and Modeling, George Mason University, and
Department of Computational and Data Sciences, George Mason University,
 Fairfax, Virginia, United States}
 \date{\today}

\begin{abstract}
 Acrylamide, with PubChem identifier CID=6579 is broadcasted to have four stable conformers contrasting with several journal publications characterizing only two or three. In this revision summary the discrepancy is clarified. Through very high precision density functional theory (DFT) calculations, three stable conformers and the three transition state barriers existing between them are verified to exist and validated with our own DFT calculations The most stable conformer is a planar molecular structure termed "sys" or "trans" in the literature. Meanwhile, a less stable structure termed "skew" pertains to two 3-dimensional structures that are energy-degenerate, but differ in their structure for being mirrored images of each other. Vibrational spectra, partial atomic charges, Cartesian coordinates, and Intrinsic Reaction Coordinate paths are summarized and recalculated with DFT  at the {$\omega$B97XD/Def2TZVPP} level for the three stable acrylamide isomers: the sys/trans lowest in energy structure, and the two skew mirrored structures.\\

{\bf{keywords}} acrylamide, Density Functional Theory, DFT, Intrinsic Reaction Coordinate, IRC, IR spectrum.
\end{abstract}

\maketitle

Over time, acrylamide (CH$_2$=CH-CONH$_2$) \cite {APA} has been studied from density functional theory (DFT) for the investigation of several molecular properties. In 2000, the B3LYP/6-311+G** and B3LYP/cc-pVTZ study \cite{acrylMDFREQ} reported two stable acrylamide conformers termed  \textit{syn} and \textit{skew} identifying the transition state between them. The \textit{syn} conformer had the lowest electronic energy and its infrared (IR) vibrational spectrum was provided \cite{acrylMDFREQ}. These two structures differed approximately by a $160^o$ rotation about the C-C bond connecting the C=C bonded carbons to the amide radical, with the \textit{syn} being planar and the \textit{skew} being a 3D structure. Later in 2005 \cite{acrylMD}, calculations based on B3LYP/6-31G* including the dispersion correction (gd3) \cite{Grimme} reported two planar, optimized conformers labeled \textit{trans} and \textit{cis}. While the \textit{trans} structure and its vibrational spectrum were similar to those previously reported \cite{acrylMDFREQ}, the \textit{cis} structure was planar, hence differing from the reported \textit{skew} 3D structure \cite{acrylMDFREQ}. Most recently in 2022 \cite{acrylMD2}, calculations at the B3LYP/aug-cc-pVTZ level reported two optimized stable conformers labeled \textit{syn} and \textit{anti}, the former possessing the lowest energy. In 2023, a new calculation \cite{acrylFREQ} reproduced the previous work \cite{acrylMD2} based on further validation bringing closer the calculated and experimental IR spectrum values. From this literature, the emerging scenario establishes existence of two stable conformers, with the \textit{syn}/\textit{trans} stability and structure being certainly planar and well documented, while the higher energy conformer termed \textit{skew} \cite{acrylMDFREQ}, \textit{cis} \cite{acrylMD}, or \textit{anti} \cite{acrylMD2} has conflicting and vague reports concerning its structure and stability. In addition, the PubChem \cite{APA} database provides four acrylamide conformer geometries, a planar one (\textit{conformer 1}) identifiable as the \textit{syn, trans} \cite{acrylMDFREQ,acrylMD} structure, a second planar structure (\textit{conformer 2}) possibly identifiable as the \textit{cis, anti} \cite{acrylMD,acrylMD2} structure, and two 3D mirrored-structures (\textit{conformers 3, 4}) with the amide group rotated $90^o$ out of the plane.

The goal of this recalculation study is asserting with high precision DFT calculations the acrylamide stable conformer structures corresponding to minima of the electronic energy surface (PES), providing their Cartesian coordinates and vibrational spectra, and determining the PES energy barriers required for transitioning between the different stable conformers. In this review study we confirm with higher precision: (i) there are three stable acrylamide conformers that have distict minima in the PES, (ii) the three PES transition state saddle points (energy barriers) between them, and (iii) the Intrinsic Reaction Coordinate (IRC) \cite{hratchian_IRC,hratchian_IRCa} that traces the path from each PES minimum, across the transition state, toward the landing minimum. 

The selected DFT methodoloy is {$\omega$B97XD/Def2TZVPP} \cite{wB97XD,Def2TZVPP}, which was systematically compared to  
{B3LYP/6-311+G**} \cite{B3LYP,6-311+G} high accuracy calculations. The Gaussian 16 \cite{gaussian} package with its structural optimization \cite{Berny} was used throughout, employing {\it opt=verytight}, {\it Int=Ultrafine,} and {\it scf=QC} for results reported therein. The positional root mean square deviation ({RMSD}) between any two acrylamide structures was calculated via {VMD} \cite{VMD} and its plugin \cite{RMSD}.  The latter aligns the compared structures by optimizing consecutive rotations between specified groups of atoms.

The structural optimization process started by investigating one by one the four acrylamide geometries provided by {PubChem} \cite{APA} and optimizing each of them within the two DFT approaches. Optimization of \textit{conformers 1,3,4} resulted in three distinct minima of the potential energy surface (PES) termed S1, S2, S3, depicted in Fig. 1. The S1 stable isomer is planar with a dihedral angle \cite{gaussview} $\theta=180^o$ between the C=C and C=O bonds, while S2 and S3 are mirrored 3-dimensional stable isomers displaying $\theta=28.15^o$ ($\omega$B97XD) or $\theta=26.15^o$. Additionally, Figure 1 depicts the transition state structures calculated within the $\omega$B97XD approach as discussed in upcoming paragraphs.
\begin{figure}[htbp]  
\centering \begin{subfigure}[b]{0.48\textwidth}
        \centering
        \includegraphics[width=\textwidth]{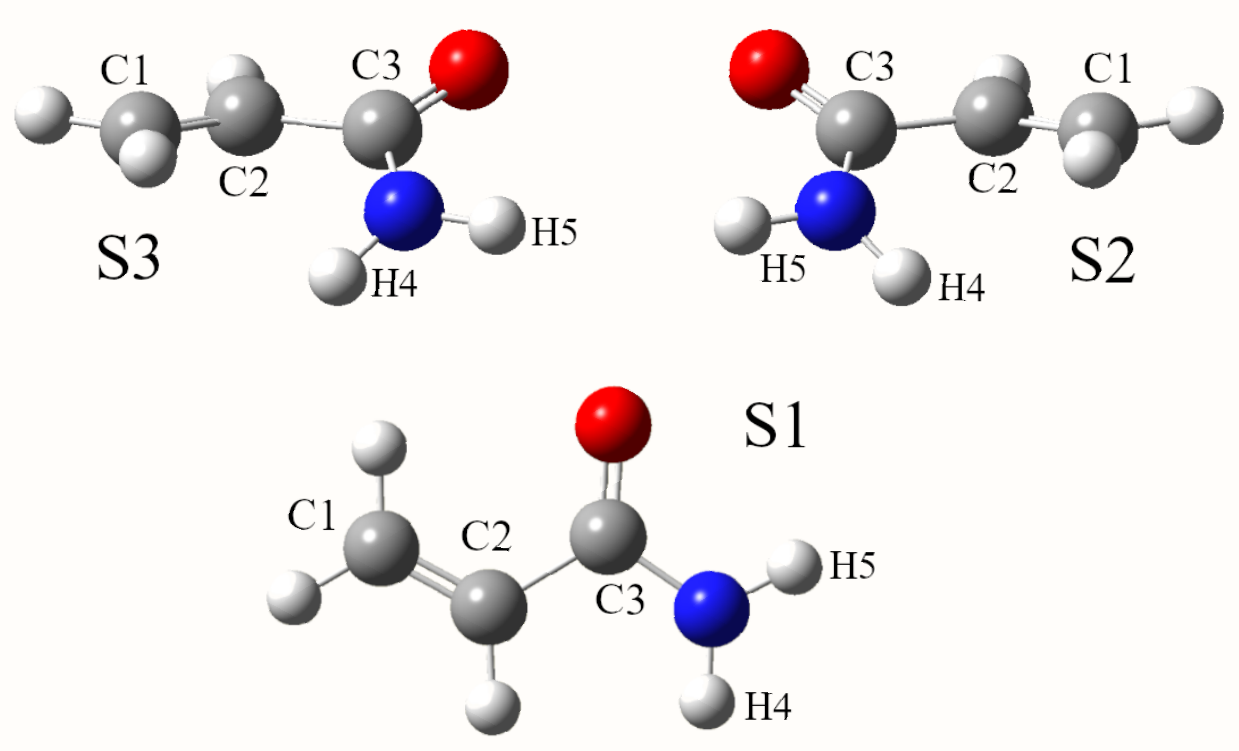}
        \caption{Stable isomers}
        \label{fig:1_left}
    \end{subfigure}
    \hfill 
    \begin{subfigure}[b]{0.48\textwidth}
        \centering
        \includegraphics[width=0.9\textwidth]{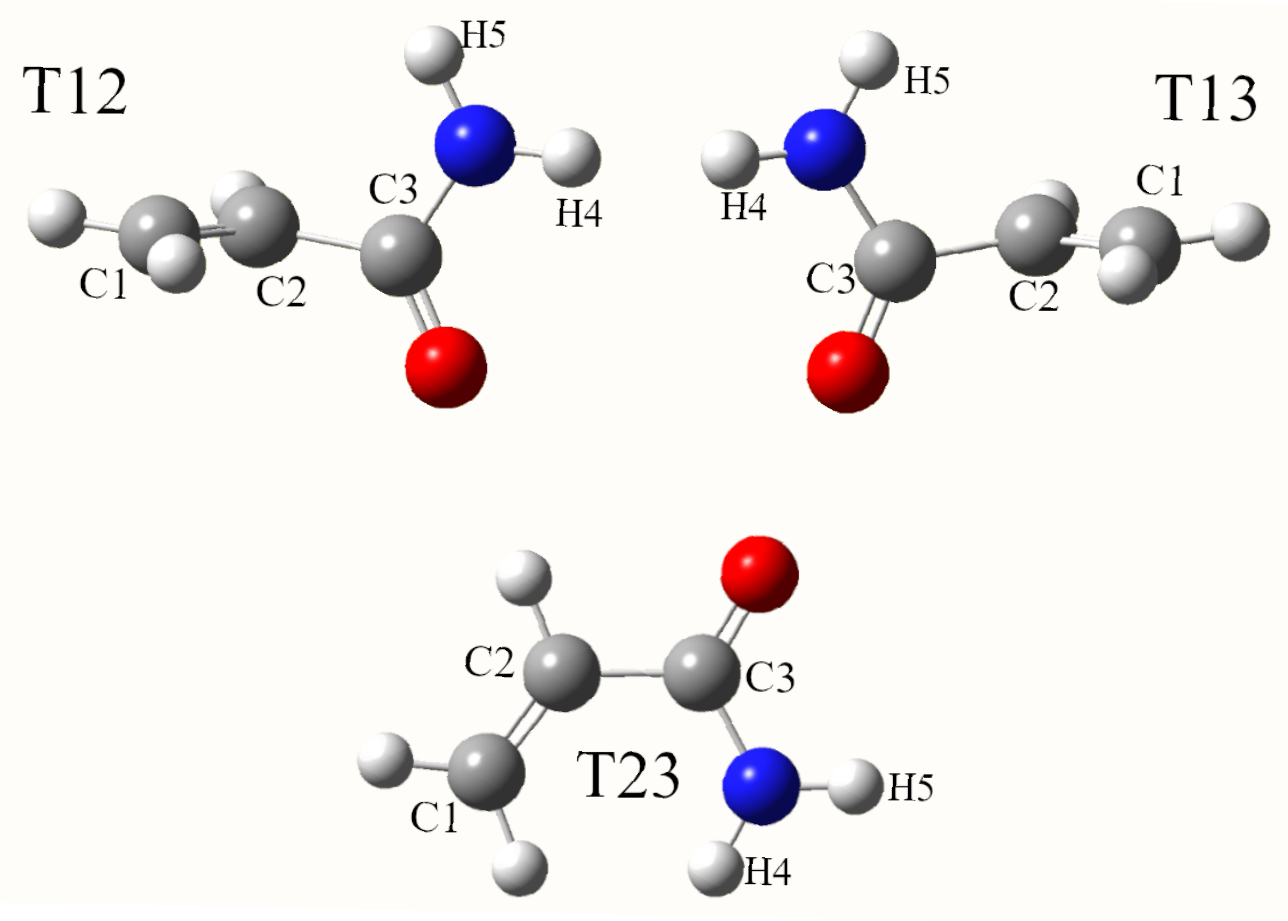}
        \caption{Transition state structures}
        \label{fig:1_right}
    \end{subfigure}
\caption{Visual depiction of the three optimized acrylamide stable conformer structures S1, S2, S3 (left) and the three calculated transition state structures between them, T12, T13, T23 (right). Calculations are done within the $\omega$B97XD/Def2TZVPP approach.}
\end{figure}
\noindent (B3LYP). It is confirmed that the  ground state of the three stable conformers are singlet states. Table I lists their electronic energies {$^1$E}, the zero point energy {E$_0$}, and the triplet state energies {$^3$E} at the same geometry of the singlets. Table II provides the vibrational spectra of the S1, S2, S3 stable conformers confirming positive frequencies that validate the optimized structures. However, the PubChem \cite{APA} \textit{conformer 2} structure did optimize to a PES first order saddle point (transition state) and is termed T23.
\begin{table}[h!] 
\centering
\caption{The $\omega$B97XD/Def2TZVPP and B3LYP/6-311+G** ground state electronic energy ($^{1}E$) of the acrylamide conformers $S1, S2, S3$, the zero point energy (E$_0$), and the triplet states electronic energy ($^{3}E$). $\Delta E$ is the difference between each conformer's ground state energy and that of S1. }
\label{tab:tab1}
\centering
\footnotesize
\begin{tabular}{lccccc} 
\hline
\centering
 & \multicolumn{4}{c}{$\omega$B97XD/Def2TZVPP} \\ 
\multicolumn{1}{c}{Isomer} & \multicolumn{1}{c}{$^1$E (Ha)} & \multicolumn{1}{c}{$\Delta$ E (kcal/mol)} & \multicolumn{1}{c}{E$_0$ (Ha)} & \multicolumn{1}{c}{$^3$E (Ha)}\\
 \hline
$S1$ & -247.317051 &     0.0    & 0.079667 & -247.151662 \\
$S2$ & -247.315108 & 1.219251 & 0.079923 & -247.155982 \\
$S3$ & -247.315108 & 1.219251 & 0.079923 & -247.155983 \\
$T12$ & -247.310170 & 4.317894 &       -     &   -   \\
$T13$ & -247.310170 & 4.317894 &       -     &   -   \\
$T23$ & -247.314477 & 1.614582 &       -     &   -   \\
\hline
& \multicolumn{4}{c}{B3LYP/6-311+G${**}$}  \\ 
\multicolumn{1}{c}{Isomer} & \multicolumn{1}{c}{E (Ha)} & \multicolumn{1}{c}{$\Delta$ E (kcal/mol)} & \multicolumn{1}{c}{E$_0$ (Ha)} & \multicolumn{1}{c}{E$_T$ (Ha)} \\
\hline
$S1$ & -247.377398 &     0.0    & 0.078479 & -247.2315563  \\
$S2$ & -247.375148 & 1.411897 & 0.078812 & -247.2286759  \\
$S3$ & -247.375148 & 1.411897 & 0.078812 & -247.2286761  \\
$T12$ & -247.370225 & 4.501126 &    -    & -  \\
$T13$ & -247.370215 & 4.507402 &    -    & -  \\
$T23$ & -247.374670 & 1.711587 &    -    & -  \\
\hline
\end{tabular}
\end{table}
\begin{table}[h!]
\centering
\caption{IR spectra of the acrylamide S1 and S2/S3 from $\omega$B97XD/Def2TZVPP and B3LYP/6-311+G**. Frequencies are scaled \cite{CCCBDB} by a factor of 0.9552 for the former and 0.967 for the latter approach.}
\label{tab:tab2}
\footnotesize
\setlength{\tabcolsep}{1.6 pt}.
\begin{tabular}  {cccc||cccc}   
\hline
\multicolumn{4}{c}{S1} & \multicolumn{4}{c}{S2/S3} \\
\multicolumn{2}{c}{$\omega$B97XD} & \multicolumn{2}{c}{B3LYP} & \multicolumn{2}{c}{$\omega$B97XD} & \multicolumn{2}{c}{B3LYP} \\
\hline
$\nu(cm^{-1})$ &Intensity&$\nu(cm^{-1})$ &Intensity&$\nu(cm^{-1})$ &Intensity&$\nu(cm^{-1})$ &Intensity\\
\hline
 99.03 & 21.79& 99.48 & 51.52          & 98.19 & 2.86& 91.28 & 3.52  \\
176.91 & 171.99 &145.19 &162.44     & 267.45 & 24.49&  265.56 & 18.28 \\
 270.68 & 7.31 &267.02 & 7.41          & 325.46 & 164.94& 329.45 & 177.84  \\
451.93 & 10.18 & 449.59 & 11.87       & 407.86 & 13.89& 411.70 & 15.52  \\
 453.39 & 4.81 &451.64 & 4.85          & 508.31 & 8.73& 507.54 & 8.43  \\
591.02 & 2.77&586.21 & 2.26           & 541.91 & 11.82& 537.40 & 15.22  \\
596.20 & 10.41&596.42 & 9.01         & 581.19 & 6.15& 572.48 & 6.69  \\
785.00 & 5.99&785.07 & 5.79            & 791.66 & 16.68& 787.22 & 20.41 \\
795.40 & 21.03&790.07 & 20.13       & 795.39 & 10.33& 792.95 & 5.54  \\
977.24 & 14.33&970.09 & 4.24         & 956.77 & 42.20& 946.34 & 42.19  \\
982.49 & 37.95&977.55 & 52.20       & 998.51 & 15.28& 995.67 & 21.52  \\
1001.23 & 4.15&1,001.36 & 6.29      & 1010.64 & 15.30& 1,007.23 & 17.66  \\
1075.69 & 2.00&1,073.55 & 3.33      & 1071.45 & 2.06& 1,074.56 & 2.27  \\
1244.92 & 91.03&1,241.67 & 122.36      & 1257.93 & 0.88& 1,263.94 & 3.28  \\
1311.46 & 49.14&1,306.85 & 47.95        & 1317.73 & 107.42& 1,309.81 & 127.73  \\
1386.01 & 102.06&1,390.02 & 82.68     & 1398.26 & 34.26 & 1,402.75 & 26.02  \\
1557.65 & 135.61&1,566.10 & 120.93    & 1555.51 & 109.16& 1,565.14 & 91.85  \\
1647.75 & 36.76 &1,627.66 & 48.18       & 1639.18 & 44.89& 1,622.47 & 54.86  \\
1722.66 & 290.09&1,698.10 & 300.07    & 1722.17 & 346.36& 1,692.51 & 351.22  \\
3022.42 & 5.89 &3,033.82 & 7.78           & 3019.36 & 4.19& 3,031.81 & 4.64  \\
3039.13 & 11.00&3,045.05 & 11.98         & 3064.90 & 0.38& 3,072.80 & 0.65  \\
3119.10 & 1.13&3,128.43 & 1.95             & 3105.26 & 5.90& 3,112.40 & 8.44  \\
 3482.56 & 54.95&3,476.12 & 50.15       & 3474.84 & 40.61& 3,468.96 & 35.94  \\
3613.45 & 46.46&3,604.37 & 44.32        & 3600.63 & 41.26& 3,591.07 & 38.28 \\
\hline
\end{tabular}
\end{table}

Despite S2 and S3 possessing an equal value of $^1$E (Table I), their 3D structures are different as validated by the RMSD between them. Indeed, each pair taken from the S1, S2, S3 stable isomers yielded RMSD of 1.305 \AA\ for (S1-S2) or (S1-S3), 0.546 \AA\ for (S2-S3) within the $\omega$B97XD approach and 2.067 \AA\ for (S1-S2) or (S1-S3), 0.517 \AA\ for (S2-S3) within the B3LYP approach. As reported in Table II,  there is significant difference between the S1 vibrational spectrum and the S2 or S3 identical spectra. In addition, the partial atomic charges are not equal, as listed in Table III. The three acrylamide stable conformers are polar molecules. The dipole moment of S1 is in the molecular plane with a magnitude of 3.6512 D, while the dipole moment of the 3D conformers S2 and S3 are equal in magnitude, 3.9083 D, with direction opposite to each other. For completeness, Table IV provides the Cartesian coordinates of the three stable acrylamide conformers in their ground state. 

\begin{table}[h!] 
\centering
\caption{Partial atomic charges in $e$ of the S1, S2, S3  acrylamide stable isomers in their ground state from the Mulliken and ESP\cite{singh,ESP} population analyses within the $\omega$B97XD/Def2TZVPP approach.}
\label{tab:tab3}
\footnotesize
\begin{tabular}{lccc||ccc}
\hline
 & \multicolumn{3}{c}{Mulliken charges} & \multicolumn{3}{c}{ESP charges} \\
\hline
Atom & \multicolumn{1}{l}{S1} & \multicolumn{1}{l}{S2} & \multicolumn{1}{l}{S3} & \multicolumn{1}{l}{S1} & \multicolumn{1}{l}{S2} & \multicolumn{1}{l}{S3} \\
\hline
H1 & 0.11 & 0.11 & 0.11 & 0.16 & 0.17 & 0.17 \\
H2 & 0.12 & 0.09 & 0.09 & 0.16 & 0.17 & 0.17 \\
C1 & -0.16 & -0.19 & -0.19 & -0.26 & -0.32 & -0.32 \\
H3 & 0.08 & 0.12 & 0.12 & 0.16 & 0.16 & 0.15 \\
C2 & -0.17 & -0.14 & -0.14 & -0.32 & -0.23 & -0.22 \\
C3 & 0.35 & 0.33 & 0.33 & 0.81 & 0.71 & 0.69 \\
O & -0.39 & -0.38 & -0.38 & -0.57 & -0.56 & -0.55 \\
N & -0.23 & -0.24 & -0.24 & -1.00 & -0.83 & -0.82 \\
H4 & 0.14 & 0.14 & 0.14 & 0.41 & 0.35 & 0.35 \\
H5 & 0.16 & 0.16 & 0.16 & 0.43 & 0.39 & 0.39 \\
\hline
Sum & 0.0 & 0.0 & 0.0 & 0.0 & 0.0 & 0.0\\
\hline
\end{tabular}
\end{table}

\begin{table}[ht!]
\centering
\caption{Cartesian coordinates (in \AA) of the acrylamide S1, S2, S3 conformers optimized within both, the $\omega$B97XD/Def2TZVPP and the B3LYP/6-311+G** approaches.}
\label{tab:tab4}
\footnotesize
 \begin{tabular}{c|ccc|ccc|ccc}
\hline
 \multicolumn{10}{c}{$\omega$B97XD/Def2TZVPP} \\
 \hline
     & \multicolumn{3}{c}{S1} & \multicolumn{3}{c}{S2} & \multicolumn{3}{c}{S3} \\
Atom  & X  & Y & Z & X  & Y & Z & X  & Y & Z \\
\hline  
H1 & -2.90 & -0.57 & 0.00 & 2.92 & -0.30 & -0.04  & -2.92 &-0.30 & -0.04 \\
H2 & -2.00 &  1.06 & 0.00 & 1.86 &  1.12 & -0.55  & -1.86 & 1.12 & -0.55 \\
C1 & -1.97 & -0.02 & 0.00 & 1.92 &  0.12 & -0.14  & -1.92 & 0.12 & -0.14 \\
H3 & -0.75 & -1.73 & 0.00 & 0.94 & -1.60 &  0.54  & -0.94 & -1.60 & 0.54 \\
C2 & -0.80 & -0.65 & 0.00 & 0.85 & -0.58 &  0.19  & -0.85 & -0.58 & 0.19 \\
C3 & 0.48 & 0.13  & 0.00  &-0.56  &-0.13 & 0.03   & 0.56 & -0.13 & 0.03 \\
O  & 0.52 & 1.34  & 0.00  & -1.46 & -0.94 & -0.14 & 1.46 & -0.94 & -0.14 \\
N  & 1.60 & -0.64  & 0.00 & -0.78 &  1.21 &  0.05 &  0.78 &  1.21 &  0.05 \\
H4 & 1.57 & -1.64  & 0.00 & -0.10 &  1.84 & 0.42  &  0.10 &  1.84 & 0.42 \\
H5 & 2.49 & -0.18  & 0.00 & -1.74 &  1.52 & 0.01  &  1.74 &  1.52 & 0.01 \\
\hline
 \multicolumn{10}{c} {B3LYP/6-311+G**}\\
\hline
H1 &  -2.91 & -0.58 & 0.00  & -0.10 & 1.85 & 0.42   & -0.10 & -1.86 & 0.42 \\
H2 &  -2.02 & 1.06 & 0.00   & -1.75 & 1.52 & 0.01   & -1.75 & -1.52 & 0.01 \\
C1 &  -1.98 & -0.03 & 0.00  & 1.94 & 0.12 & -0.13   & 1.94 & -0.12 & -0.13 \\
H3 &  -0.74 & -1.73 & 0.00  & 0.93 & -1.62 & 0.50   & 0.93 & 1.62 & 0.50 \\
C2 &  -0.80 & -0.65 & 0.00  & 0.85 & -0.59 & 0.18   & 0.85 & 0.59 & 0.18 \\
C3 &  0.48 & 0.13 & 0.00    & -0.56 & -0.13 & 0.02  & -0.56 & 0.13 & 0.02 \\
O  &  0.52 & 1.35 & 0.00    & -1.47 & -0.94 & -0.13 & -1.47 & 0.94 & -0.13 \\
N  &  1.61 & -0.64 & 0.00   & -0.79 & 1.22 & 0.04   & -0.79 & -1.22 & 0.04 \\
H4 &  1.58 & -1.65 & 0.00   & 2.93 & -0.31 & -0.03  & 2.93 & 0.31 & -0.03 \\
H5 &  2.50 & -0.18 & 0.00   & 1.88 & 1.13 & -0.51   & 1.88 & -1.13 & -0.51\\

\hline
\end{tabular}
\end{table}

Determination of the transition state structures T12 between S1 and S2, T13 between S1 and S3, and T23 between S2 and S3 were achieved through the optimization approach QST3 \cite{gaussian,QST3}. The optimized PES points found correspond to three first order saddles with electronic energies reported in Table I and geometries displayed in Figure 1. As expected, the T12 and T13 are mirrored transition structures with the same energy barrier arising between S1 and S2 than between S1 and S3. The T12 and T13 transition structures differ by a RMSD less than 0.086 \AA\ from the two 3D structures provided by PubChem \cite{APA} (\textit{conformers 3, 4}). Meanwhile, the T23 optimized transition structure is planar and a result of employing the PubChem \cite{APA} \textit{conformer 2} as input to the optimization process. A RMSD of 0.12 \AA\ exists between T23 and PubChem's \textit{conformer 2}. Figure 2 (left) depicts the IRC from the S1 minimum of the PES, climbing the energy barrier associated to S12 (or S13) and landing to the S2 (or S3) PES minimum. In addition, Figure 2 (right) illustrates the IRC from the S3 minimum transitioning over the S23 energy barrier and reaching the S2 minimum of the PES. The latter IRC is an interesting validation of how one of the two mirrored conformers, either S2 or S3 can transition into the other once the energy barrier is overcome. The RMSD between the three saddle points molecular structures verify their differences: 1.404 \AA\ or 1.408 \AA\ for (T12-T13) within the $\omega$B97XD and B3LYP approaches, respectively, and 0.850 \AA\ (T12-T23) or (T13-T23) within both DFT approaches. \\

\begin{figure}[h!]
\centering
    \includegraphics[width=0.45\linewidth]{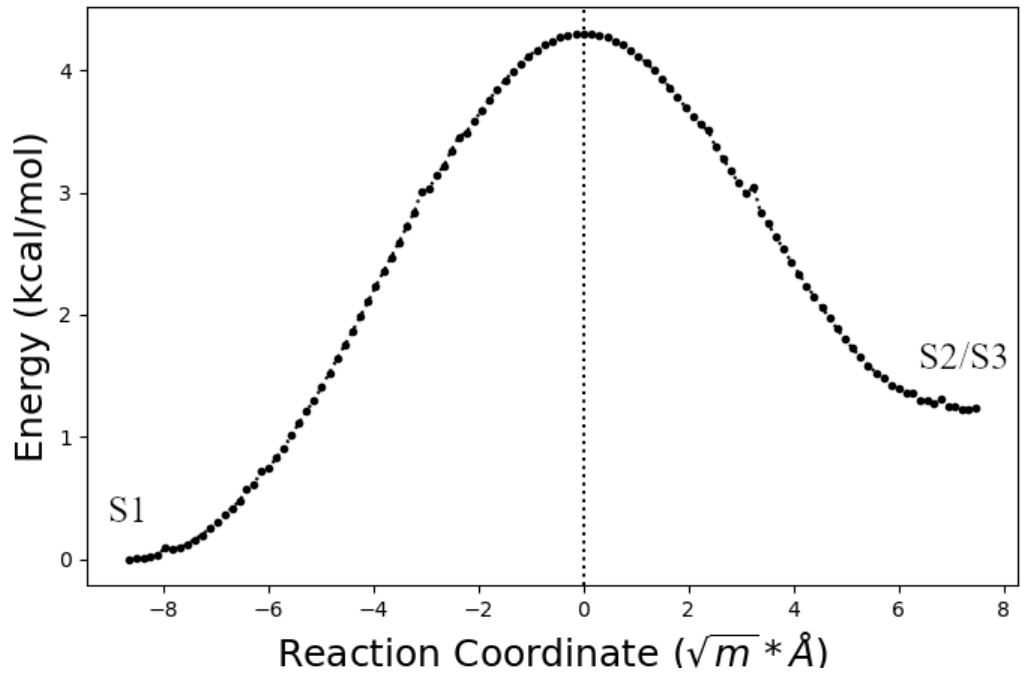}
    \includegraphics[width=0.47\linewidth]{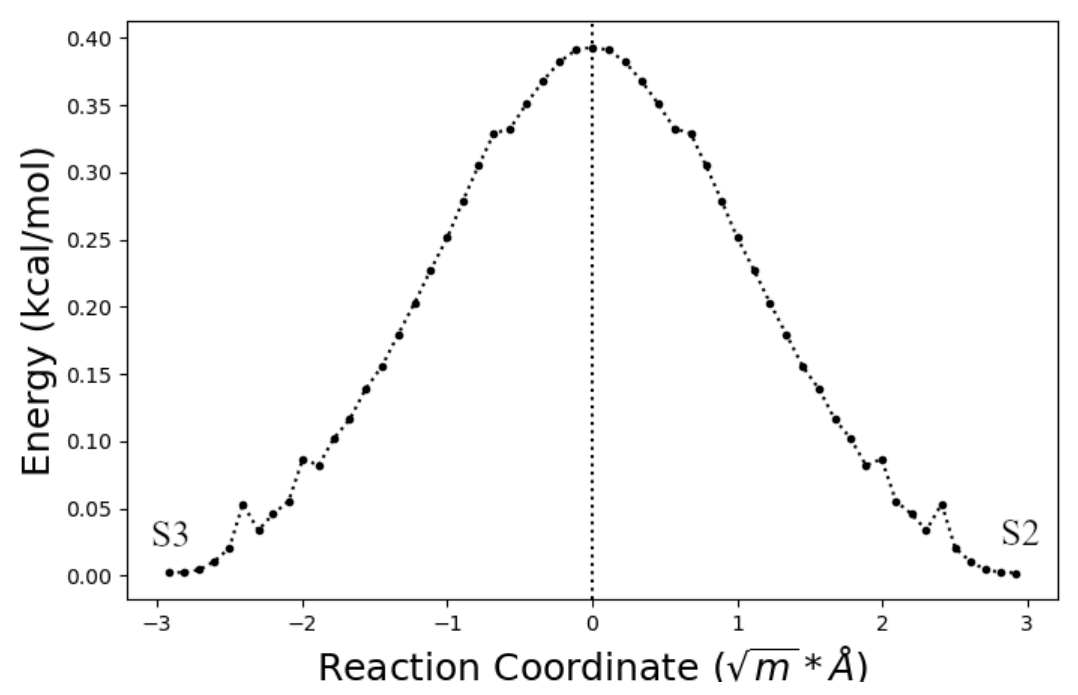}
    \caption{The IRC within the $\omega$B97XD/Def2TZVPP approach. Left: the S1 conformer overcoming the transition state barrier T12 (or T13) and reaching the S2 (or S3) conformer.  Right:  the S2 (or S3) conformer overcoming the T23 energy barrier and reaching the S3 (or S2) conformer. IRC step size is 0.052918 \AA.}
    \label{fig:fig2}
\end{figure}

In conclusion, from the DFT perspective, acrylamide has three stable isomers S1, S2, and S3. The lowest in energy isomer is S1, a singlet state with planar structure termed in the literature \textit{syn, trans, conformer 1}\cite{acrylMDFREQ,acrylMD,acrylMD2,APA}. Depending upon the DFT approach, the two out-of-plane, mirrored-structure, isomers S2 and S3 are 3-dimensional geometries displaying a dihedral angle $\theta=28.15-26.15^o$ and having the same ground state energy of 1.22-1.41 kcal/mol above S1. Either S2 or S3 corresponds to the structure termed \textit{skew} in the literature\cite{acrylMDFREQ}. Despite the energy proximity between S1 and the two mirrored isomers S2, S3, the vibrational frequency spectra are different and should be easily identifiable experimentally. The triplet states of the three stable isomers lay more than 90 kcal/mol above the singlets. The well defined transition state structures T12 and T13 are the PubChem broadly distributed geometries termed \textit{conformers 3, 4,}\cite{APA} while the well defined transition state T23 with planar structure was referred to in the literature as a stable conformer termed \textit{cis, anti, conformer 2}\cite{acrylMD,acrylMD2,APA}.\\

{\bf{Acknowledgments}}
We acknowledge the Office of Research Computing of George Mason University for the computer time allocated to this project.

{\footnotesize 
\bibliography{Scott_Blaisten-Barojas_acrylamide.bib} 

\begin{thebibliography}{21}%
\makeatletter
\providecommand \@ifxundefined [1]{%
 \@ifx{#1\undefined}
}%
\providecommand \@ifnum [1]{%
 \ifnum #1\expandafter \@firstoftwo
 \else \expandafter \@secondoftwo
 \fi
}%
\providecommand \@ifx [1]{%
 \ifx #1\expandafter \@firstoftwo
 \else \expandafter \@secondoftwo
 \fi
}%
\providecommand \natexlab [1]{#1}%
\providecommand \enquote  [1]{``#1''}%
\providecommand \bibnamefont  [1]{#1}%
\providecommand \bibfnamefont [1]{#1}%
\providecommand \citenamefont [1]{#1}%
\providecommand \href@noop [0]{\@secondoftwo}%
\providecommand \href [0]{\begingroup \@sanitize@url \@href}%
\providecommand \@href[1]{\@@startlink{#1}\@@href}%
\providecommand \@@href[1]{\endgroup#1\@@endlink}%
\providecommand \@sanitize@url [0]{\catcode `\\12\catcode `\$12\catcode
  `\&12\catcode `\#12\catcode `\^12\catcode `\_12\catcode `\%12\relax}%
\providecommand \@@startlink[1]{}%
\providecommand \@@endlink[0]{}%
\providecommand \url  [0]{\begingroup\@sanitize@url \@url }%
\providecommand \@url [1]{\endgroup\@href {#1}{\urlprefix }}%
\providecommand \urlprefix  [0]{URL }%
\providecommand \Eprint [0]{\href }%
\providecommand \doibase [0]{https://doi.org/}%
\providecommand \selectlanguage [0]{\@gobble}%
\providecommand \bibinfo  [0]{\@secondoftwo}%
\providecommand \bibfield  [0]{\@secondoftwo}%
\providecommand \translation [1]{[#1]}%
\providecommand \BibitemOpen [0]{}%
\providecommand \bibitemStop [0]{}%
\providecommand \bibitemNoStop [0]{.\EOS\space}%
\providecommand \EOS [0]{\spacefactor3000\relax}%
\providecommand \BibitemShut  [1]{\csname bibitem#1\endcsname}%
\let\auto@bib@innerbib\@empty
\bibitem [{APA(2025)}]{APA}%
  \BibitemOpen
  \href {https://pubchem.ncbi.nlm.nih.gov/compound/Acrylamide} {\bibinfo
  {title} {{C}hem compound summary for {CID} 6579, {A}crylamide}} (\bibinfo
  {year} {2025}),\ \bibinfo {note} {last accessed December 5, 2025}\BibitemShut
  {NoStop}%
\bibitem [{\citenamefont {Marstokk}\ \emph {et~al.}(2000)\citenamefont
  {Marstokk}, \citenamefont {M{\"o}llendal},\ and\ \citenamefont
  {Samdal}}]{acrylMDFREQ}%
  \BibitemOpen
  \bibfield  {author} {\bibinfo {author} {\bibfnamefont {K.~M.}\ \bibnamefont
  {Marstokk}}, \bibinfo {author} {\bibfnamefont {H.}~\bibnamefont
  {M{\"o}llendal}},\ and\ \bibinfo {author} {\bibfnamefont {S.}~\bibnamefont
  {Samdal}},\ }\href {https://doi.org/10.1016/S0022-2860(99)00362-2} {\bibfield
   {journal} {\bibinfo  {journal} {J. Mol. Struct.}\ }\textbf {\bibinfo
  {volume} {524}},\ \bibinfo {pages} {69} (\bibinfo {year} {2000})}\BibitemShut
  {NoStop}%
\bibitem [{\citenamefont {Duarte}\ \emph {et~al.}(2005)\citenamefont {Duarte},
  \citenamefont {da~Costa},\ and\ \citenamefont {Amado}}]{acrylMD}%
  \BibitemOpen
  \bibfield  {author} {\bibinfo {author} {\bibfnamefont {A.}~\bibnamefont
  {Duarte}}, \bibinfo {author} {\bibfnamefont {A.~A.}\ \bibnamefont
  {da~Costa}},\ and\ \bibinfo {author} {\bibfnamefont {A.}~\bibnamefont
  {Amado}},\ }\href@noop {} {\bibfield  {journal} {\bibinfo  {journal} {J.
  Molec. Struct.: {THEOCHEM}}\ }\textbf {\bibinfo {volume} {723}},\ \bibinfo
  {pages} {63} (\bibinfo {year} {2005})}\BibitemShut {NoStop}%
\bibitem [{\citenamefont {Grimme}(2006)}]{Grimme}%
  \BibitemOpen
  \bibfield  {author} {\bibinfo {author} {\bibfnamefont {S.}~\bibnamefont
  {Grimme}},\ }\href {https://doi.org/10.1002/jcc.20495} {\bibfield  {journal}
  {\bibinfo  {journal} {J. Comp. Chem.}\ }\textbf {\bibinfo {volume} {27}},\
  \bibinfo {pages} {1787} (\bibinfo {year} {2006})}\BibitemShut {NoStop}%
\bibitem [{\citenamefont {Evangelisti}\ \emph {et~al.}(2022)\citenamefont
  {Evangelisti}, \citenamefont {Melandri}, \citenamefont {Negri}, \citenamefont
  {Coreno}, \citenamefont {Prince},\ and\ \citenamefont {Maris}}]{acrylMD2}%
  \BibitemOpen
  \bibfield  {author} {\bibinfo {author} {\bibfnamefont {L.}~\bibnamefont
  {Evangelisti}}, \bibinfo {author} {\bibfnamefont {S.}~\bibnamefont
  {Melandri}}, \bibinfo {author} {\bibfnamefont {F.}~\bibnamefont {Negri}},
  \bibinfo {author} {\bibfnamefont {M.}~\bibnamefont {Coreno}}, \bibinfo
  {author} {\bibfnamefont {K.}~\bibnamefont {Prince}},\ and\ \bibinfo {author}
  {\bibfnamefont {A.}~\bibnamefont {Maris}},\ }\href
  {https://doi.org/10.3390/photochem2030032} {\bibfield  {journal} {\bibinfo
  {journal} {Photochem}\ }\textbf {\bibinfo {volume} {2}},\ \bibinfo {pages}
  {463–478} (\bibinfo {year} {2022})}\BibitemShut {NoStop}%
\bibitem [{\citenamefont {Chong}(2023)}]{acrylFREQ}%
  \BibitemOpen
  \bibfield  {author} {\bibinfo {author} {\bibfnamefont {D.~P.}\ \bibnamefont
  {Chong}},\ }\href {https://doi.org/10.1016/j.elspec.2023.147359} {\bibfield
  {journal} {\bibinfo  {journal} {J. Electron Spectrosc. Relat. Phenom.}\
  }\textbf {\bibinfo {volume} {266}},\ \bibinfo {pages} {147359} (\bibinfo
  {year} {2023})}\BibitemShut {NoStop}%
\bibitem [{\citenamefont {Hratchian}\ and\ \citenamefont
  {Schlegel}(2004)}]{hratchian_IRC}%
  \BibitemOpen
  \bibfield  {author} {\bibinfo {author} {\bibfnamefont {H.~P.}\ \bibnamefont
  {Hratchian}}\ and\ \bibinfo {author} {\bibfnamefont {H.~B.}\ \bibnamefont
  {Schlegel}},\ }\href@noop {} {\bibfield  {journal} {\bibinfo  {journal} {J.
  Chem. Phys.}\ }\textbf {\bibinfo {volume} {120}},\ \bibinfo {pages} {9918}
  (\bibinfo {year} {2004})}\BibitemShut {NoStop}%
\bibitem [{\citenamefont {Hratchian}\ and\ \citenamefont
  {Schlegel}(2005)}]{hratchian_IRCa}%
  \BibitemOpen
  \bibfield  {author} {\bibinfo {author} {\bibfnamefont {H.~P.}\ \bibnamefont
  {Hratchian}}\ and\ \bibinfo {author} {\bibfnamefont {H.~B.}\ \bibnamefont
  {Schlegel}},\ }\href {https://doi.org/10.1021/ct0499783} {\bibfield
  {journal} {\bibinfo  {journal} {J. Chem. Theory Comput.}\ }\textbf {\bibinfo
  {volume} {1}},\ \bibinfo {pages} {61} (\bibinfo {year} {2005})}\BibitemShut
  {NoStop}%
\bibitem [{\citenamefont {Chai}\ and\ \citenamefont
  {Head-Gordon}(2008)}]{wB97XD}%
  \BibitemOpen
  \bibfield  {author} {\bibinfo {author} {\bibfnamefont {J.~D.}\ \bibnamefont
  {Chai}}\ and\ \bibinfo {author} {\bibfnamefont {M.}~\bibnamefont
  {Head-Gordon}},\ }\href {https://doi.org/10.1039/B810189B} {\bibfield
  {journal} {\bibinfo  {journal} {Phys. Chem. Chem. Phys.}\ }\textbf {\bibinfo
  {volume} {10}},\ \bibinfo {pages} {6615} (\bibinfo {year}
  {2008})}\BibitemShut {NoStop}%
\bibitem [{\citenamefont {Weigend}\ and\ \citenamefont
  {Ahlrichs}(2005)}]{Def2TZVPP}%
  \BibitemOpen
  \bibfield  {author} {\bibinfo {author} {\bibfnamefont {F.}~\bibnamefont
  {Weigend}}\ and\ \bibinfo {author} {\bibfnamefont {R.}~\bibnamefont
  {Ahlrichs}},\ }\href {https://doi.org/10.1039/B508541A} {\bibfield  {journal}
  {\bibinfo  {journal} {Phys. Chem. Chem. Phys.}\ }\textbf {\bibinfo {volume}
  {7}},\ \bibinfo {pages} {3297} (\bibinfo {year} {2005})}\BibitemShut
  {NoStop}%
\bibitem [{\citenamefont {Becke}(1993)}]{B3LYP}%
  \BibitemOpen
  \bibfield  {author} {\bibinfo {author} {\bibfnamefont {A.~D.}\ \bibnamefont
  {Becke}},\ }\href {https://doi.org/10.1063/1.464913} {\bibfield  {journal}
  {\bibinfo  {journal} {J. Chem. Phys.}\ }\textbf {\bibinfo {volume} {98}},\
  \bibinfo {pages} {5648} (\bibinfo {year} {1993})}\BibitemShut {NoStop}%
\bibitem [{\citenamefont {McLean}\ and\ \citenamefont
  {Chandler}(1980)}]{6-311+G}%
  \BibitemOpen
  \bibfield  {author} {\bibinfo {author} {\bibfnamefont {A.~D.}\ \bibnamefont
  {McLean}}\ and\ \bibinfo {author} {\bibfnamefont {G.~S.}\ \bibnamefont
  {Chandler}},\ }\href {https://doi.org/10.1063/1.438980} {\bibfield  {journal}
  {\bibinfo  {journal} {J. Chem. Phys.}\ }\textbf {\bibinfo {volume} {72}},\
  \bibinfo {pages} {5639} (\bibinfo {year} {1980})}\BibitemShut {NoStop}%
\bibitem [{\citenamefont {Frisch}\ \emph {et~al.}(2016)\citenamefont {Frisch},
  \citenamefont {Trucks}, \citenamefont {Schlegel}, \citenamefont {Scuseria},\
  and\ \citenamefont {{et al.}}}]{gaussian}%
  \BibitemOpen
  \bibfield  {author} {\bibinfo {author} {\bibfnamefont {M.~J.}\ \bibnamefont
  {Frisch}}, \bibinfo {author} {\bibfnamefont {G.~W.}\ \bibnamefont {Trucks}},
  \bibinfo {author} {\bibfnamefont {H.~B.}\ \bibnamefont {Schlegel}}, \bibinfo
  {author} {\bibfnamefont {G.~E.}\ \bibnamefont {Scuseria}},\ and\ \bibinfo
  {author} {\bibnamefont {{et al.}}},\ }\href@noop {} {\emph {\bibinfo {title}
  {Gaussian}}},\ \bibinfo {address} {Wallingford CT} (\bibinfo {year}
  {2016})\BibitemShut {NoStop}%
\bibitem [{\citenamefont {Li}\ and\ \citenamefont {Frisch}(2006)}]{Berny}%
  \BibitemOpen
  \bibfield  {author} {\bibinfo {author} {\bibfnamefont {X.}~\bibnamefont
  {Li}}\ and\ \bibinfo {author} {\bibfnamefont {M.~J.}\ \bibnamefont
  {Frisch}},\ }\href {https://doi.org/10.1021/ct050275a} {\bibfield  {journal}
  {\bibinfo  {journal} {Chem. Theory and Comput.}\ }\textbf {\bibinfo {volume}
  {2}},\ \bibinfo {pages} {835} (\bibinfo {year} {2006})}\BibitemShut {NoStop}%
\bibitem [{\citenamefont {Humphrey}\ \emph {et~al.}(1996)\citenamefont
  {Humphrey}, \citenamefont {Dalke},\ and\ \citenamefont {Schulten}}]{VMD}%
  \BibitemOpen
  \bibfield  {author} {\bibinfo {author} {\bibfnamefont {W.}~\bibnamefont
  {Humphrey}}, \bibinfo {author} {\bibfnamefont {A.}~\bibnamefont {Dalke}},\
  and\ \bibinfo {author} {\bibfnamefont {K.}~\bibnamefont {Schulten}},\ }\href
  {https://doi.org/10.1016/0263-7855(96)00018-5} {\bibfield  {journal}
  {\bibinfo  {journal} {J. Mol. Graphics}\ }\textbf {\bibinfo {volume} {14}},\
  \bibinfo {pages} {33} (\bibinfo {year} {1996})}\BibitemShut {NoStop}%
\bibitem [{\citenamefont {Eargle}\ \emph {et~al.}(2006)\citenamefont {Eargle},
  \citenamefont {Wright},\ and\ \citenamefont {Luthey-Schulten}}]{RMSD}%
  \BibitemOpen
  \bibfield  {author} {\bibinfo {author} {\bibfnamefont {J.}~\bibnamefont
  {Eargle}}, \bibinfo {author} {\bibfnamefont {D.}~\bibnamefont {Wright}},\
  and\ \bibinfo {author} {\bibfnamefont {Z.}~\bibnamefont {Luthey-Schulten}},\
  }\href {https://doi.org/10.1093/bioinformatics/bti825} {\bibfield  {journal}
  {\bibinfo  {journal} {Bioinformatics}\ }\textbf {\bibinfo {volume} {22}},\
  \bibinfo {pages} {504} (\bibinfo {year} {2006})}\BibitemShut {NoStop}%
\bibitem [{\citenamefont {Dennington}\ \emph {et~al.}(2019)\citenamefont
  {Dennington}, \citenamefont {Keith},\ and\ \citenamefont
  {Millam}}]{gaussview}%
  \BibitemOpen
  \bibfield  {author} {\bibinfo {author} {\bibfnamefont {R.}~\bibnamefont
  {Dennington}}, \bibinfo {author} {\bibfnamefont {T.~A.}\ \bibnamefont
  {Keith}},\ and\ \bibinfo {author} {\bibfnamefont {J.~M.}\ \bibnamefont
  {Millam}},\ }\href@noop {} {\bibinfo {title} {Gauss{V}iew version {6}}}
  (\bibinfo {year} {2019}),\ \bibinfo {note} {{S}emichem {I}nc. {S}hawnee
  {M}ission {KS}}\BibitemShut {NoStop}%
\bibitem [{CCC(2025)}]{CCCBDB}%
  \BibitemOpen
  \href {https://cccbdb.nist.gov/vsfx.asp} {\bibinfo {title} {{CCCBDB}
  collection of vibrational frequency scaling factors}} (\bibinfo {year}
  {2025}),\ \bibinfo {note} {last accessed, December 5, 2025}\BibitemShut
  {NoStop}%
\bibitem [{\citenamefont {Singh}\ and\ \citenamefont {Kollman}(1984)}]{singh}%
  \BibitemOpen
  \bibfield  {author} {\bibinfo {author} {\bibfnamefont {U.~C.}\ \bibnamefont
  {Singh}}\ and\ \bibinfo {author} {\bibfnamefont {P.~A.}\ \bibnamefont
  {Kollman}},\ }\href@noop {} {\bibfield  {journal} {\bibinfo  {journal} {J.
  Comput. Chem.}\ }\textbf {\bibinfo {volume} {5}},\ \bibinfo {pages} {129}
  (\bibinfo {year} {1984})}\BibitemShut {NoStop}%
\bibitem [{\citenamefont {Besler}\ \emph {et~al.}(1990)\citenamefont {Besler},
  \citenamefont {Jr.},\ and\ \citenamefont {Kollman}}]{ESP}%
  \BibitemOpen
  \bibfield  {author} {\bibinfo {author} {\bibfnamefont {B.~H.}\ \bibnamefont
  {Besler}}, \bibinfo {author} {\bibfnamefont {K.~M.~M.}\ \bibnamefont {Jr.}},\
  and\ \bibinfo {author} {\bibfnamefont {P.~A.}\ \bibnamefont {Kollman}},\
  }\href {https://doi.org/10.1002/jcc.540110404} {\bibfield  {journal}
  {\bibinfo  {journal} {J. Comput. Chem.}\ }\textbf {\bibinfo {volume} {11}},\
  \bibinfo {pages} {431} (\bibinfo {year} {1990})}\BibitemShut {NoStop}%
\bibitem [{\citenamefont {Peng}\ \emph {et~al.}(1996)\citenamefont {Peng},
  \citenamefont {Ayala}, \citenamefont {Schlegel},\ and\ \citenamefont
  {Frisch}}]{QST3}%
  \BibitemOpen
  \bibfield  {author} {\bibinfo {author} {\bibfnamefont {C.}~\bibnamefont
  {Peng}}, \bibinfo {author} {\bibfnamefont {P.~Y.}\ \bibnamefont {Ayala}},
  \bibinfo {author} {\bibfnamefont {H.~B.}\ \bibnamefont {Schlegel}},\ and\
  \bibinfo {author} {\bibfnamefont {M.~J.}\ \bibnamefont {Frisch}},\ }\href
  {https://doi.org/10.1002/(SICI)1096-987X(19960115)17:1<49::AID-JCC5>3.0.CO;2-0}
  {\bibfield  {journal} {\bibinfo  {journal} {J. Comp. Chem.}\ }\textbf
  {\bibinfo {volume} {17}},\ \bibinfo {pages} {49} (\bibinfo {year}
  {1996})}\BibitemShut {NoStop}%
\end{thebibliography}%

\end{document}

\typeout{get arXiv to do 4 passes: Label(s) may have changed. Rerun}